\documentclass[aps,pre,groupedaddress,showpacs,floatfix]{revtex4}
\usepackage{graphicx}
\usepackage{dcolumn}
\usepackage{bm}
\usepackage{amssymb}
\usepackage{amsmath}
\usepackage{epsfig}

\begin{document}

\title{Nonequilibrium patterns and shape fluctuations in reactive membranes}

\author{Ramon Reigada$^{1},$ Javier Buceta$^{2}$ and Katja Lindenberg$^{3} $}
\affiliation{$^{(1)}$Departament de Qu\'{\i}mica-F\'{\i}sica, Universitat de 
Barcelona,
Avda. Diagonal 647, 08028 Barcelona, Spain\\
$^{(2)}$Parc Cient\'{\i}fic de Barcelona, Centre de Recerca en
Qu\'{\i}mica Te\'orica (CeRQT), Campus Diagonal - Universitat de
Barcelona, Edifici Modular, C/Josep Samitier 1-5, 08028 Barcelona,
Spain\\
$^{(3)}$Department of Chemistry and Biochemistry 0340, and Institute for
Nonlinear Science, University of California, San Diego,
9500 Gilman Drive, San Diego, CA 92093, USA}

\begin{abstract}
A simple kinetic model of a two-component deformable and reactive bilayer
is presented.  The two differently shaped components are interconverted
by a nonequilibrium reaction, and a phenomenological coupling between
local composition and curvature is proposed.
When the two components are not miscible, linear stability
analysis predicts, and numerical simulations show, the formation of
stationary nonequilibrium composition/curvature patterns
whose typical size is determined by the reactive process.
For miscible components, a
linearization of the dynamic equations is performed in order to evaluate
the correlation function for shape fluctuations from which the behavior
of these systems in micropipet aspiration experiments can be predicted.
\end{abstract}

\pacs{87.16.Dg, 07.05.Tp, 82.45.Mp}
\maketitle

\section{Introduction}
\label{intro}

Polar lipids self-assemble and orient, with the hydrophilic
portions facing water. The
water may be sandwiched by two lipid layers, as in the black spots in
soap bubbles that develop at locations where the two layers are closer
than the shortest wavelength of the visible spectrum and that occur just
before the bubble bursts, or the water may be outside of the two
lipid layers, as in
biomembranes. While it is known that the structure of a biological
membrane is far more complex than a simple lipid
bilayer~\cite{lip,lipnat} because of
embedded proteins, cholesterol molecules, and ions, to name just a few
of the components that provide the full functionality to this structure
that is crucial to the life of the cell, it is nevertheless acknowledged
that the lipid bilayer is the basic structural unit of \textit{all} cell
membranes.  An understanding of this simpler system is therefore
extremely important in an effort to shed light on the properties and
functioning of active transport, signaling, and adhesion in cells.
Furthermore, artificial lipid bilayers are widely used in a number of
nanotechnological applications ranging from solar energy transduction
and biosensors to drug development.  

A lipid bilayer is highly flexible and liquid-like (as is a real membrane). 
It can therefore not be viewed as a static inert boundary
but must be recognized as a dynamical structure~\cite{hous}.  In
the case of a
biological membrane, lipid bilayers serve as quasi-two-dimensional
solvents for proteins and all other components, and are intimately
involved in many biochemical processes.  Moreover, its ability to change its
shape (curvature) is an essential property of the lipid bilayer since the
formation of vesicles and the permeability properties of the cell depend
on it.

From the modeling viewpoint, it became recognized during the 90's
that some internal degrees
of freedom are necessary to understand the large variety
of conformational changes found in cell and synthetic membranes.
In particular, the local composition
of the bilayer can crucially affect its local curvature, and some models
were developed on the basis of this idea~\cite{and,sun,goz,dyn3,dyn2,dyn1}.
However, these approaches considered membranes as
equilibrium systems. In a biological context, this hypothesis is at
best hopeful.

Nonequilibrium conditions are ubiquitous in nature, and for this
reason the out-of-equilibrium behavior of membranes, both in cellular
systems and in laboratory-prepared vesicles,
has attracted much attention over the past few years.
The first successful approach to the study of nonequilibrium membranes
was introduced by J. Prost \textit{et al.}~\cite{prost}.
Roughly summarizing their modeling approach, they suggested that some
externally activated components (intramembrane proteins) act as ``pumps,''
generating forces on the membrane that locally change its curvature.
Variations, improvements and sequels of the initial
model~\cite{prost2,prost3,sum,chen} as well as
related experimental studies~\cite{prost3,prost4}
to a large extent complete the understanding of the nonequilibrium
behavior of bilayers with inserted active components.

Our interest in this work lies in a different
nonequilibrium that may arise from externally activated chemical
processes involving the transformation of the elementary membrane
lipid components.
As an example, nice experiments by P. G. Petrov \textit{et al.}~\cite{petrov}
show the effect of an ongoing photocontroled chemical reaction
on the curvature of synthetic giant vesicles. Other evidence of
chemically-induced
shape transformations is found in the nervous synaptic process:
a reaction that interconverts two differently shaped constituent
phospholipids of the membrane plays a crucial role
in the fission of vesicles in nervous cells~\cite{scales,schmidt}.
Recent experiments also point to the importance of high-curvature
lipids in many cell processes such as membrane fusion~\cite{sci04}.
So far, no models for this kind of nonequilibrium situation can be
found in the literature.  Our aim is to gain physical insight into the
role of a generic nonequilibrium reaction
acting on a membrane that is, in turn, described through a
curvature/composition coupling.  We show how the reactive process
endows the membrane with a rich pattern formation phenomenology and
specific characteristics of its shape fluctuations. 

\begin{figure}[h!]
\vspace{-0.1in}
\begin{center}
\includegraphics[width=7cm ]{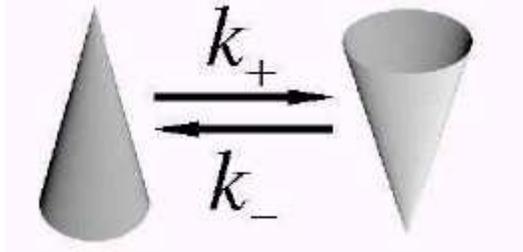}
\caption{The two membrane components $A$ (left) and $B$ (right) have
opposite cone shapes.  We arbitrarily choose a positive curvature for
$A$ and a negative one for $B$. A reaction interconverts both species with
$k_+$ and $k_-$ being the forward and backward reaction rates, respectively.}
\label{figmodel0}
\end{center}
\end{figure}

We present a simple kinetic model of a two-component
deformable bilayer with a chemical reaction. We model
nonactive membranes where the nonequilibrium nature of the system is
caused by a chemical reaction
that interconverts two differently shaped components of the bilayer
rather than by inserted active components.
In the model, the membrane is composed of two species:
lipid $A$, which is assumed to be coned-shaped, and lipid $B$, which has
an inverted cone shape (see Fig.~\ref{figmodel0}).
We consider the simplest scenario where the outer layer of the membrane
is composed of $A$ and $B$ lipids, whereas the inner layer is composed
of a single component without any curvature effect.
We also prescribe that lipids do not move between the inner and outer layers.
Both of these simplifications are made in most membrane models.
According to these assumptions, the membrane is simply modeled as a
laterally heterogeneous
elastic surface with an internal composition
order parameter locally coupled to the curvature.

Within this framework, we are basically interested in two situations.
On the one hand, for immiscible components, we show that
phase separation in the membrane
leads to the spontaneous development
of structures involving heterogeneous distributions of
both composition and curvature that
finally result in stationary finite-sized nonequilibrium domains.
This instability is studied by performing a linear stability analysis
of the kinetic equations, and the pattern-selection role of the
reactive process is established. Corresponding numerical simulations
will be shown to support these predictions.
On the other hand, for miscible components we derive
the correlation function for the membrane shape fluctuations. The fluctuation
spectrum is mainly determined by the bending energy at small surface
tension, and
here we find the same expression for nonreactive and reactive cases,
but with different bending rigidities. This change in the behavior
of the fluctuation spectrum between
equilibrium and nonequilibrium states might be easily controlled and
observed in micropipet aspiration experiments.

This paper is organized as follows. In Sec.~\ref{unstable} the
study of membranes of immiscible components is
addressed. We propose a free energy functional, derive the kinetic
equations, and perform the linear stability analysis of these
equations. Numerical simulations are carried out and
some representative results are shown.
In Sec.~\ref{stable} bilayers of miscible components are analyzed and the change of
the height fluctuation spectrum between the reactive and nonreactive
situations is established.
We conclude with a brief summary in Sec.~\ref{concl}.

\section{Membranes of immiscible components}
\label{unstable}

\subsection{Model and analytical results}
\label{modan}

The membrane is defined as a two dimensional surface with a
concentration difference
order parameter $\phi$ and a local extrinsic curvature $H$.
The rigidity of the membrane
leads to an elastic energy contribution
$\frac{\kappa}{2}\int{\left(H-H_{sp}(\phi)\right)^2} dxdy$
to the total energy,
where $\kappa$ is the bending rigidity
modulus, and the spontaneous (equilibrium) curvature
$H_{sp}(\phi)$ reflects the shape asymmetry between the two lipid
components.  For simplicity, we adopt a linear dependence on $\phi$,
$H_{sp}=\phi H_{0}$, with $H_0 > 0$ according to the schematic
in Fig.~\ref{figmodel0}.  In the Monge parametrization~\cite{geom}
a deformable surface is described by $(x,y,h(x,y))$, where $h(x,y)$ is
the displacement (height)
field for the local separation from the flat conformation. This
representation is valid for surfaces that are nearly flat with
only gradual variations of $h$, and allows the approximation
$H \approx \nabla^2 h$.
As a function of these variables, the proposed energy functional reads
\begin{equation}
{\mathcal{F}}= \int \left[ -\frac{\alpha}{2} \phi^{2}+ \frac{\beta}{4} \phi^{4}+ \frac{\gamma}{2}
|{\mathbf{\nabla}} \phi| ^{2} + \frac{\kappa}{2}
( \nabla^2 h - \phi H_0)^2  \right] dx dy,
\label{fgen}
\end{equation}
where the first three terms correspond to the typical
Ginzburg-Landau expansion responsible for phase separation
($\alpha$, $\beta$, $\gamma > 0$), with
an equilibrium concentration difference
$\phi_{eq}=\pm \sqrt{\alpha / \beta}$, and a typical
interface length $\zeta = \sqrt{\gamma/\alpha}$.
For self-assembled free membranes, the surface tension contribution
($\frac{\sigma}{2}|{\mathbf{\nabla}} h| ^{2}$) in the free energy
can be neglected, and we have not included it in Eq. (\ref{fgen}).

The kinetics of $\phi$ follows a conserved scheme~\cite{sancho} plus
the reaction contributions,
\begin{equation}
\frac{\partial \phi}{\partial t} = 
D \nabla^{2} \left[ \frac{\partial {\mathcal{F}}}{\partial \phi}
\right] - \Gamma (\phi - \phi_0),
\label{phivar}
\end{equation}
where $\Gamma=k_++k_-$ and $\phi_0=(k_--k_+)/(k_++k_-)$.
$k_+$ and $k_-$ are the forward and backward reaction rate constants,
respectively.
Considering a permeable membrane (i.e., ignoring hydrodynamic interactions)
we adopt the following relaxational equation for the
evolution of the height field, 
\begin{equation}
\frac{\partial h }{\partial t}=- \Lambda \frac{\delta {\mathcal{F}}}{\delta
h },
\label{hvar}
\end{equation}
where $\Lambda$ is a mobility parameter proportional to the inverse of
the typical relaxation time $\tau_h$.

\begin{figure}[h!]
\begin{center}
\includegraphics[width=6cm ]{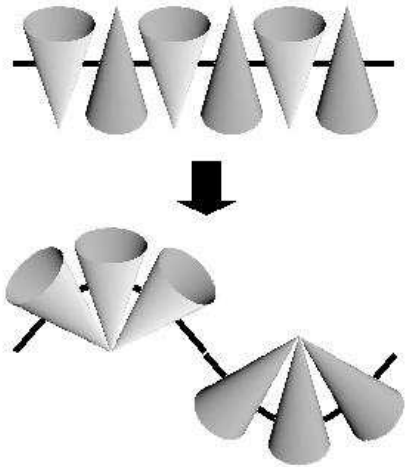}
\caption{Schematic representation of the composition/curvature coupling
effect in an unstable membrane.}
\label{figmodel}
\end{center}
\end{figure}

The kinetic equations can readily be adimensionalized: energy is
measured in units of $k_BT$, time
in units of $\tau_h$, and length in units of $\sqrt{D \tau_h}$.
In terms of the new dimensionless
parameters, the kinetic equations become
\begin{eqnarray}
\frac{\partial \phi}{\partial t} &=& 
(\kappa H_0^{2} - \alpha) \nabla^{2} \phi + 3 \beta \phi^{2}\nabla^{2}\phi
+ 6 \beta \phi
|{\mathbf{\nabla}} \phi|^{2} \nonumber\\ \label{kineq}
&&~~-\gamma \nabla^{4}\phi -\kappa H_0 \nabla^{4}h
 - \Gamma \left( \phi - \phi_0 \right), \\
\frac{\partial h }{\partial t}&=&- \kappa \nabla^{4}h
+\kappa H_0 \nabla^{2} \phi \nonumber .
\end{eqnarray}
At thermal equilibrium Eqs.~(\ref{kineq})
describe the spinodal decomposition
of two immiscible components. Due to the composition/curvature
coupling $H_0$, both fields
will form complementary patterns. As phase segregation progresses,
membrane regions
with positive (negative) $\phi$ deform in such a way that the
curvature become positive (negative),
as shown in Fig.~\ref{figmodel}.
In the absence of reaction, this coarsening process does
not end until there is complete segregation
into two large domains. A nonequilibrium reaction such as the one we
propose converts one species into the other.  This amounts to
a large-scale mixing mechanism that counteracts the short-scale
ordering effect of phase separation. Therefore, the segregated
structures grow only until mixing and ordering effects compensate,
resulting in a stationary pattern.
These types of nonequilibrium patterns are also found in other systems
such as polymer blends~\cite{patsel,tcong1} as well as in
monomolecular adsorbtion on metal surfaces~\cite{dewell,mexPRE},
and have to be distinguished from typical Turing patterns~\cite{turing}.
Even though they emerge from the same kind of instablity
(see below), Turing patterns are expected in systems with species
with different diffusivities. In the model presented here,
the domains result from the competition between a local 
thermodynamic affinity of equal species and a nonequilibrium
reaction mixing effect. As we will see,
linear stability analysis and numerical simulations support this idea. 

The stationary uniform state corresponds to
$\overline{\phi}=\phi_0$ and arbitrary $\overline{h}$.
The linear stability of these uniform solutions is tested by adding
small plane-wave perturbations of wave numer $q$ and
linearizing Eqs.~(\ref{kineq}).
This procedure determines the $2\times 2$ linearization
matrix ${\mathcal{L}}$, with the following coefficients,
\begin{align}
\begin{split}
{\mathcal{L}}_{11} &= -q^2
\left[ \left( \kappa H_0^2 -\alpha +3 \beta \phi_0^2 \right)
+\gamma q^2  \right] - \Gamma \\
{\mathcal{L}}_{12} &= -\kappa H_0 q^4\\
{\mathcal{L}}_{21} &= -\kappa H_0 q^2 \\
{\mathcal{L}}_{22} &= -\kappa q^4.
\label{coefl}
\end{split}
\end{align}

The eigenvalues $\omega_q$ of the Jacobian associated with the
matrix ${\mathcal{L}}$ correspond to the linear growth rates
of the perturbations.
Solving the eigenvalue problem we obtain
$\omega_q=\frac{1}{2}\left(Tr[{\mathcal{L}}]
\pm \sqrt{\Delta[{\mathcal{L}}]}\right)$,
where $\Delta[{\mathcal{L}}]=Tr[{\mathcal{L}}]^2
-4Det[{\mathcal{L}}]$.
At the instability boundary,
$Re(\omega_q)$ vanishes for one finite wave number that is defined
as the first unstable mode.
If the imaginary part of $\omega_q$ is not zero at this wave number,
we have a wave bifurcation. The condition
for this bifurcation is obtained by requiring $Tr[{\mathcal{L}}]=0$
and $\Delta[{\mathcal{L}}]<0$.  For positive $\kappa$, $\Gamma$,
$\alpha$, $\beta$ and $\gamma$, these conditions do not apply at
any real wave number, and consequently, wave instability is not found in
this model.

\begin{figure}[h!]
\begin{center}
\includegraphics[width=8cm ]{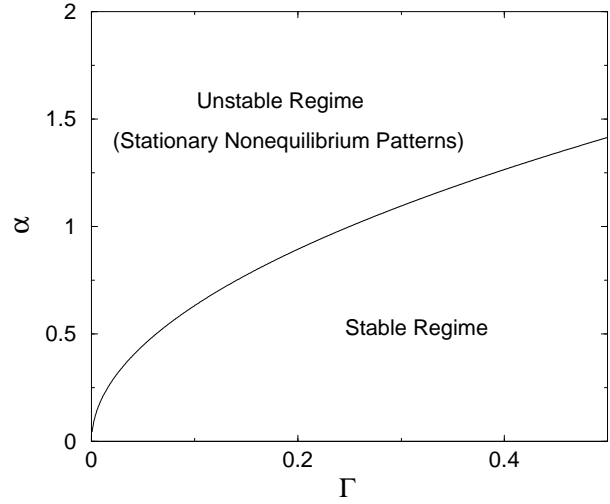}
\caption{Phase diagram for the case of immiscible components in the plane $(\alpha,\Gamma)$ for
$\gamma=1$ and $\phi_0=0$. The region $\alpha < 0$ (not shown) corresponds to a membrane of miscible constituents
displaying a flat shape.}
\label{figdiagfas}
\end{center}
\end{figure}

On the other hand, if the imaginary part of the growth rate is zero at the
bifurcation point, we have a Turing-like bifurcation.
The condition for this bifurcation is $Det[{\mathcal{L}}]=0$,
whose analytical expression is easily obtained, yielding
the following condition for the model parameters,
\begin{equation}
\Gamma = \Gamma_c \equiv \frac{\left( \alpha - 3 \beta \phi_{0}^2 \right)^2}{4\gamma},
\label{tbif}
\end{equation}
and for the wave vector of the first unstable mode,
\begin{equation}
q^2_c = \frac{\alpha - 3 \beta \phi_{0}^2}{2 \gamma}
=\sqrt{\frac{\Gamma_c}{\gamma}}.
\label{q0t}
\end{equation}
A phase diagram is shown in Fig.~\ref{figdiagfas}. In the unstable
region, Turing-like patterns
emerging from the competition between phase separation and reaction
are predicted.
In the stable phase, although the two components are not
inherently miscible, the reaction completely mixes
the system, and the bilayer becomes stable and essentially flat. 

\begin{figure}[h!]
\begin{center}
\includegraphics[width=8cm ]{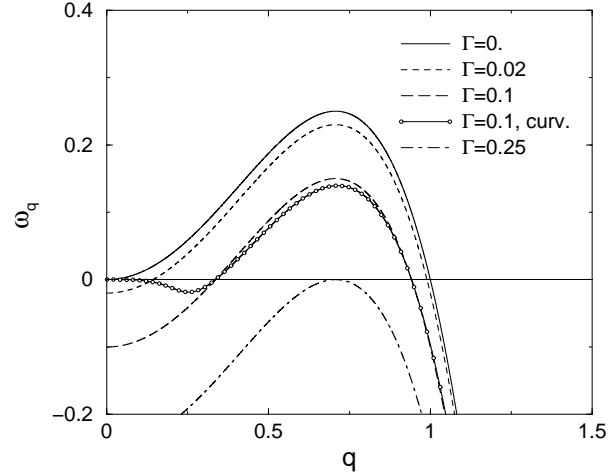}
\caption{Dispersion relation functions $\omega_q$ at different
values of $\Gamma$.
The other parameters are held fixed at $\alpha=\gamma=1$,
$\phi_0=0$ and $\kappa=H_0=0$, except
for the curve with circle symbols (labeled with `\& curv' in the legends) that corresponds
to $\kappa=10$ and $H_0=0.2$.}
\label{figomega}
\end{center}
\end{figure}

In the unstable phase, $\Gamma < \Gamma_c$, there is a range of unstable modes.
Comparing with the equilibrium case ($\Gamma=0$) where a continuous
range of modes starting at $q=0$ is unstable, reaction stabilizes
the long wavelength modes so that, as we anticipate,
the long time distribution of the system results in a
stationary nonequilibrium pattern with a finite size (see further discussion
in Sec. \ref{numres}).
Notice in Fig.~\ref{figomega} that increasing $\Gamma$ reduces
the range of unstable modes progressively until $\Gamma$
reaches the marginal value in Eq.~(\ref{tbif}), above which there is no
longer any instability. 
The limits of the range of unstable modes are given by
\begin{equation}
q^2_{\pm}=\frac{\alpha - 3 \beta \phi_{0}^2}{2\gamma}
\pm \frac{1}{2\gamma}\sqrt{(\alpha - 3 \beta \phi_{0}^2
)^2 -4\gamma \Gamma},
\label{qpm}
\end{equation}
which are independent of curvature parameters. Curvature reduces the
unstable mode growth rates (see line with circles in Fig.~\ref{figomega}) but without
changing either $q_{\pm}$ or the
marginal condition~(\ref{tbif}). The effect of curvature is
exclusively related to the kinetics of the
phase separation process (see below).

Before numerically solving the kinetic equation,
we can seed light on the expected pattern formation process by performing a weakly
nonlinear analysis using the amplitude equations technique. This
analysis allows us to compute a solution for our problem near the bifurcation
threshold, find the universality class of the pattern formation mechanism, and
explain some properties such as the spatial arrangement of the patterns
found in the numerical simulations.

For simplicity, we restrict the calculation of the amplitude equations to one
spatial dimension. In spite of this restriction, we will be able to predict the kind
of spatial arrangement found in two dimensions by using the universality
properties of the amplitude equations. Details of the derivation of the
amplitude equations are presented in the appendix. The analysis reveals that a
solution for Eqs. (\ref{kineq})
near the bifurcation to pattern formation reads%
\begin{align*}
\phi\left(  x,t\right)   &  =\phi_{0}+ \left[ A\exp\left(  iq_{c}x\right)
+c.c. \right] \text{,}\\
h\left(  x,t\right)   &  = \overline{h} + \left[ B\exp\left(  iq_{c}x\right)  +c.c. \right] \text{,}%
\end{align*}
where $c.c.$ stands for the complex conjugate,
and the amplitudes $A$ and $B$ satisfy the equations%
\begin{align}
\begin{split}
\partial_{t}A &  =\left(  \Gamma_{c}-\Gamma\right)  A\left[  1-\frac
{2\beta\left(  3\alpha+7\beta\phi_{0}^{2}\right)  }{\left(  \alpha-3\beta
\phi_{0}^{2}\right)  ^{2}}\left\vert A\right\vert ^{2}\right]  +2\left(
\alpha-3\beta\phi_{0}^{2}+\frac{H_{0}^{2}\kappa}{2}\right)  \partial
_{xx}A+\frac{3H_{0}\kappa\left(  \alpha-3\beta\phi_{0}^{2}\right)  }{\gamma
}\partial_{xx}B\text{,}\\
\partial_{t}B &  =H_{0}\kappa\partial_{xx}A+\frac{3\kappa\left(  \alpha
-3\beta\phi_{0}^{2}\right)  }{\gamma}\partial_{xx}B\text{.}
\label{ae}
\end{split}
\end{align}
Moreover, since $B$ is purely relaxational, we can adiabatically eliminate it
and reduce (\ref{ae}) to a single evolution equation for the amplitude $A$,%
\begin{align}
\begin{split}
\partial_{t}A &  =\left(  \Gamma_{c}-\Gamma\right)  A\left[  1-\frac
{2\beta\left(  3\alpha+7\beta\phi_{0}^{2}\right)  }{\left(  \alpha-3\beta
\phi_{0}^{2}\right)  ^{2}}\left\vert A\right\vert ^{2}\right]  +4\gamma
q_{c}^{2}\partial_{xx}A\text{,}\\
B &  =-\frac{H_{0}}{6q_{c}^{2}}A\text{.}
\label{gl}
\end{split}
\end{align}
Therefore, the amplitudes satisfy the real Ginzburg-Landau equation. If
$\Gamma>\Gamma_{c}$ then the amplitudes relax to zero and a homogeneous
state ($\phi=\phi_0$ and arbitrary $h=\overline{h}$)
is obtained. However, if $\Gamma<\Gamma_{c}$
then the amplitudes reach a stationary value and a pattern develops if
$q_{c}\in\mathbb{R}$. Notice that the amplitudes present a negative-positive
aspect, that is, at sites where $A$ reaches a maximum (minimum) $B$ reaches a
minimum (maximum). This is simply a consequence of the
particular selection of sign for the spontaneous curvature $H_{0}$. With regard
to the nonlinear term, $A\left\vert A\right\vert ^{2}$, note first that its
coefficient is always negative. As a result, the nonlinearity always plays a
stabilizing role, the bifurcation to pattern formation is supercritical for
all values of the parameters, and no hysteresis may occur. Secondly, the
nonlinear term provides information about the relevant modal interactions. At
this point, we can take advantage of the universality properties of the
amplitude equations to infer the spatial arrangement of the pattern in two
dimensions. Since the Swift-Hohenberg model also shares the same universality
class when performing an amplitude equation analysis, namely, it also reduces
to the real Ginzburg-Landau equation \cite{hecke,newell}, the pattern
formation mechanism is the same in both models. It is well-known in that case
that the modal interactions in two dimensions are such that if the inversion
symmetry is preserved then roll-like patterns develop. On the other hand, if
that symmetry is not fulfilled, a hexagonal structure appears. Note that in
Eqs. (\ref{kineq}) the inversion symmetry,
$\{ \phi  \rightarrow -\phi , h  \rightarrow -h \}$
is only satisfied if $\phi_{0}=0$. Thus we expect a roll-like pattern in
that case, while hexagons will develop otherwise. 

\subsection{Numerical results}
\label{numres}

Numerical integration of Eqs.~(\ref{kineq})
has been performed in two dimensions
using an explicit Euler scheme in a square lattice with periodic
boundary condition.
Small random perturbations around $\overline{\phi}=\phi_0$ and
$\overline{h}=0$ are implemented as initial conditions.
The coordinate step $\Delta x$ was chosen equal to $1$, and the
time step was usually $\Delta t=10^{-4}$ to assure good numerical accuracy
(length and time in dimensionless simulation units).
The numerical results presented in this section correspond to
highly immiscible components (deep quench,
$\alpha=1$ and $\beta=1$, leading to an equilibrium value of
$\phi_{eq}=\pm 1$), and an interface thickness of the
order of the space discretization ($\zeta=1$, which leads to $\gamma=1$).
The bending rigidity modulus is taken
equal to $10$ (in units of $k_BT$)~\cite{kapa}, and we consider
two constituent lipids of very different shapes by setting
$H_0=0.2$. All our numerical results are consistent with the
predictions of the linear stability analysis and the amplitude equations.
Representative numerical results are presented below.

In Fig.~\ref{figpat} we show the simulation results for three
different situations. The first row
corresponds to small $\Gamma=0.05$ and a critical quench
($\phi_0=0$), showing the development of
a laberynthine pattern that is still evolving at $t=2000$ (the
longest time for the snapshots
shown in the figure). The coarsening process, however, is progressively
slowed down later on.
The $\phi$-field profiles of these domains (horizontal cross-sectional
cuts through the pattern) reveal regions where $\phi_{eq}$ is $+1$ or
$-1$ connected by abrupt boundaries that indicate the short spatial
range over which the equilibrium order parameter changes from one
value to the other. This is a signature of the fact that
the phase separation process is dominant for this value of $\Gamma$.
The second and third sets of panels correspond to large 
$\Gamma=0.2$ (the marginal condition
for the selected parameters is $\Gamma_c=0.25$) showing laberynthine
(critical quench in the second row)
and quasi-hexatic droplet-like (off critical quench, $\phi_0=0.1$,
in the third set) patterns. A spatial Fourier transform of this
stationary pattern shows a clear hexagonal structure.
These domains are already stationary at times
longer than $t=2000$, and can indeed be considered as nonequilibrium
stable phases of the system, involving both composition and
curvature modulations.
Their $\phi$- and $h$-field profiles have a smooth
harmonic shape due to the strength of the reactive process.
In both cases, since we are close to the bifurcation boundary,
small local deviations from the stationary values are obtained.
We especially monitor the variations of the height field and find that
$\langle |\nabla h| \rangle < 0.05$ is satisfied, so that the model has a real
physical correspondence.  

\begin{figure}[htb]
\begin{center}
\epsffile{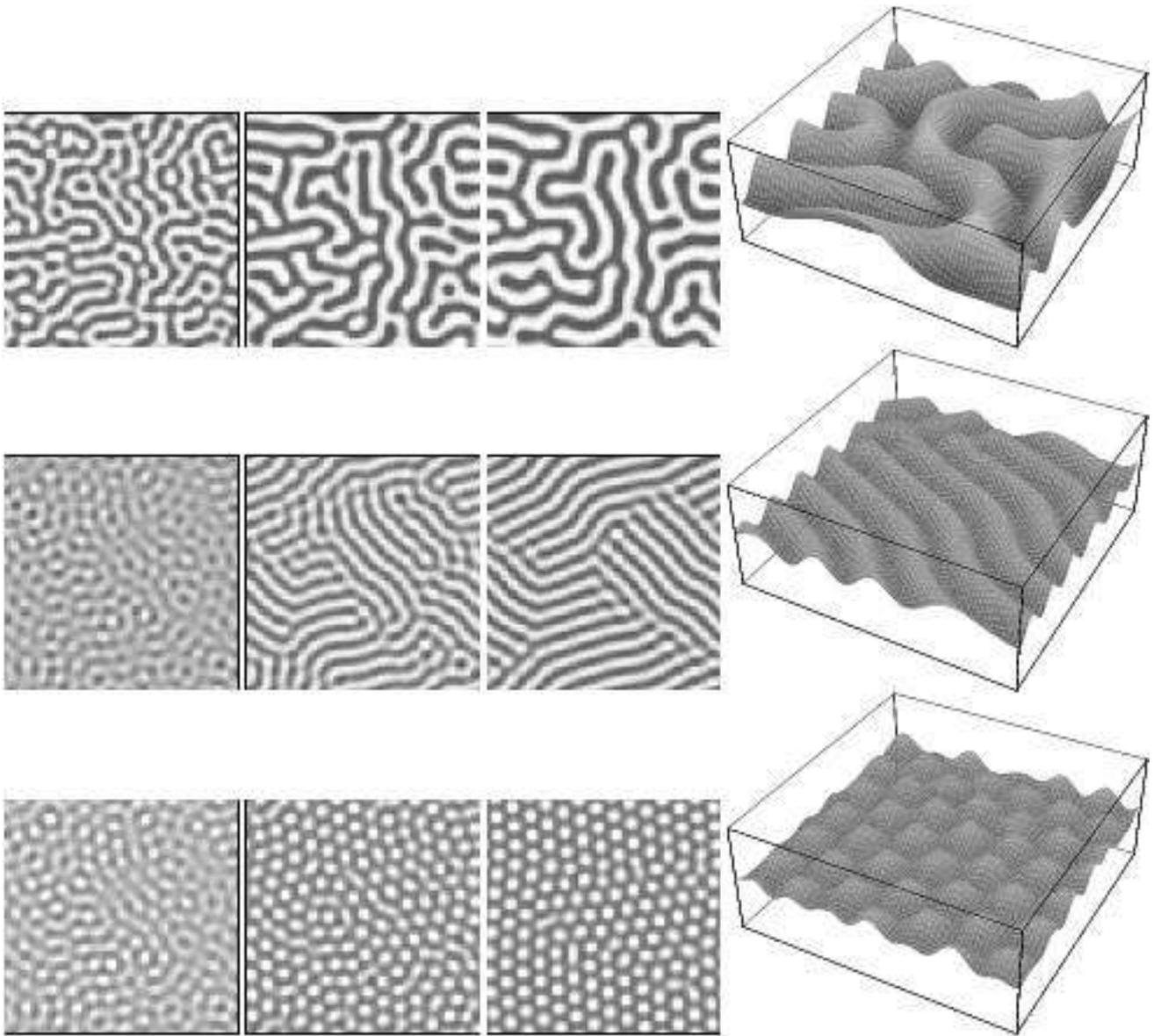}
\end{center}
\caption{Three sets of selected patterns resulting from the numerical
simulation based on Eqs.~(\ref{kineq}).
First row: $\Gamma=0.05$, $\phi_0=0$. Second row: $\Gamma=0.2$, $\phi_0=0$.
Third row: $\Gamma=0.2$, $\phi_0=0.1$ (off critical quench), showing
an array of droplet-like domains rich in the minority species.
The first three panels in each row correspond to the
$\phi$-field distributions
of a $128 \times 128$ system at $t=100$, $400$ and $2000$
(from left to right). Darker (lighter)
regions are richer in the $A$ ($B$) lipid. In the last
panel of each row the height field has been plotted in three
dimensions for a $64\times 64$ portion of the corresponding system
at $t=2000$. Some exaggeration along the vertical direction
has been applied.}
\label{figpat}
\end{figure}

In order to assess the kinetic ordering process more quantitatively,
we monitor the domain size $L(t)$
computed from the composition correlation functions
$\langle \phi({\bf r'},t) \phi({\bf r'}+{\bf r},t) \rangle$
(the same results are obtained using the height correlation functions).
Here, the brackets $\langle \cdots \rangle$
indicate not only an average over orientations of ${\bf r}$ and
over surface positions ${\bf r'}$, but also
over different realizations of random initial conditions. 
As has been reported in other studies, curvature considerably slows
down the coarsening segregation process. This is specially evident
when reaction is absent, and for this
situation different kinetic approaches have been
invoked. A mean-field kinetic scheme
by Taniguchi~\cite{dyn3} leads to extremely slow laberynthine stripe
growth obeying $L(t) \sim t^r$ with $r=0.1$, instead of the usual
spinodal decomposition growth exponent of $1/3$~\cite{lifs}. Later,
Monte Carlo simulations~\cite{dyn2,dyn1} showed
how, in the long time evolution, those stripes break up into
disconnected buds that subsequently diffuse and coalesce.
In our model, when the reactive term is removed, similar results as
those presented in Ref.~\cite{dyn3} are obtained,
and no stripe breakup is observed. The reason for the disagreement
between the analytic and Monte Carlo schemes is probably due to the fact
that continuum models subjected to a specific parametrization
of the surface such as given in Eqs.~(\ref{kineq})
do not allow for overhangs that are surely crucial in the late
stages of membrane phase segregation.

\begin{figure}[h!]
\begin{center}
\includegraphics[width=8cm ]{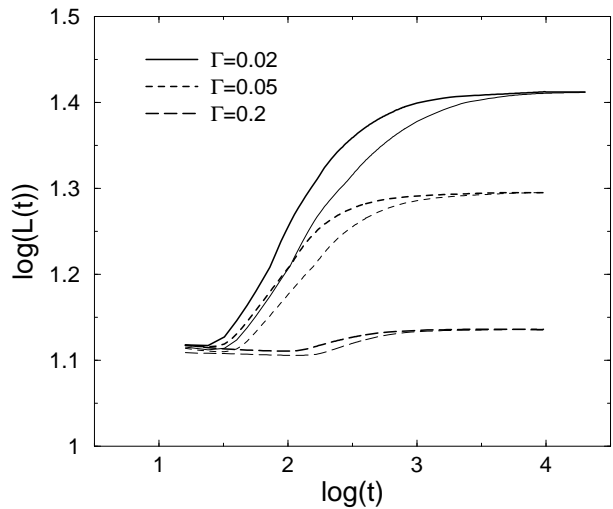}
\caption{Log-log plots of $L(t)$ vs $t$ for different values of $\Gamma$.
The values of $L(t)$ have been computed as an average over $10$ runs
of the correlation functions in
$128\times 128$ systems. In all cases $\phi_0=0$. The
curvature parameters are set to $\kappa=10$, $H_0=0.2$ for
the thin lines, and $\kappa=H_0=0$ for the thicker lines.}
\label{figlt}
\end{center}
\end{figure}

However, the cases in which there is no reaction or in which the
reaction is weak are not of interest
for us, since in these cases the system evolves
in such away that large gradients of the height displacement field
occur. As noted above, when this happens the Monge surface
parametrization is not valid and the model, although 
mathematically robust, does not describe the physical behavior
of any real system.
In the reactive cases, however, the slow kinetics is still
observed for the times prior to stationary pattern
formation. This is observed in Fig.~\ref{figlt}, where a set of
curves showing the domain size $L(t)$ is presented for several
values of $\Gamma$.
The situations with and without curvature are compared for each case.
Notice how the deformable systems
evolve more slowly than the nondeformable ones, although the same final
stationary size is achieved.
There is no theory to explain such a slowing down effect,
but some hand-waving arguments may explain the physical reasons for this
behavior. Monitoring of the different energy contributions in
Eq.~(\ref{fgen}) indicates that at early
times (when the kinetics with and without curvature are still quite
similar) interfacial energy due to the rapid
formation of composition domains is much larger than any other
energy contribution. In other words, phase separation
proceeds and membrane curvature follows the composition change.
At this stage the reduction of the interfacial energy governs the
coarsening process, leading to the $1/3$ Lifshitz-Slyozov exponent.
Later, at intermediate times, when composition domains become larger, the
height order parameter is no longer able to keep up with the
phase separation process and some curvature energy
is stored, causing the whole coarsening process to slow down. 

\begin{figure}[h!]
\begin{center}
\includegraphics[width=8cm ]{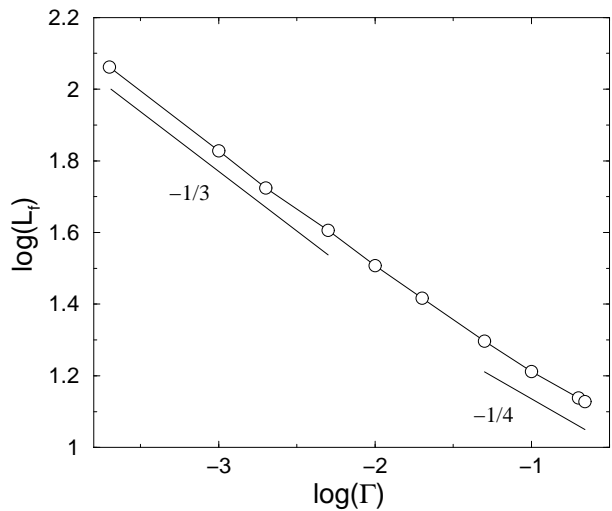}
\caption{Log-log plot of $L_f$ vs. $\Gamma$.
The slopes $1/4$ and $1/3$ are plotted for comparison with the large
and small $\Gamma$ regimes, respectively. All the points have been calculated
for $\phi_0=0$, $\kappa=10$ and $H_0=0.2$, at sufficiently long times
to consider the growth process practically halted. We have performed an
average over $10$ realizations. The system sizes are $128\times 128$
for large $\Gamma$, and $256\times 256$ for small $\Gamma$.}
\label{figlinf}
\end{center}
\end{figure}

The final size of the nonequilibrium stationary domains,
$L_f=L(t \rightarrow \infty)$, is determined by the reaction parameter.
The dependence of the final pattern size on the
reaction parameter, $L_f \sim \Gamma^{-s}$, has been
largely discussed in the literature~\cite{patsel,chri,liu}. The results
of our model also reproduce the two limiting
behaviors: $s=1/4$ for large $\Gamma$ (close to its marginal value), and
$s=1/3$ for small $\Gamma$.
In Fig.~\ref{figlinf}, $L_f$ is plotted for different values of the
reaction parameter.
Linear fits for the four first and the last four data points give
the slopes $0.323\pm 0.006$ and $0.26\pm 0.006$,
respectively. The derivation of the exponent values for the
rigid situation (nondeformable surfaces) is performed
by minimizing an effective free energy expression (where
the reactive term is included via Green's functions) in a square
(small $\Gamma$) and a harmonic (large $\Gamma$) approximation
for the stationary concentration field~\cite{chri,liu}.
In the present model, the curvature kinetics relaxationally
follows the concentration dynamics. At longer times (when the
stationary patterns are already achieved) no significant curvature
energy is present in the system,
so that we can neglect the curvature
contributions in Eq.~(\ref{fgen}) and the results in Refs.~\cite{chri,liu}
are recovered.

In Eq.~(\ref{hvar}) we ignored the hydrodynamic effects due to
the background fluid velocity
by considering a permeable membrane through which the fluid drains
freely as the system evolves.
Solvent hydrodynamic effects~\cite{seif} can be approached through
a renormalized height mobility
$\Lambda=\left( 4 \mu q \right)^{-1}$ in the second of
Eqs.~(\ref{kineq}) in Fourier space,
where $\mu$ is the solvent kinematic
viscosity. The inclusion of this $q$-dependent
mobility does not change the main results of the stability analysis
and numerical simulations shown so far.

Before ending this section, we check the applicability of our
results to real membrane systems. 
We adimensionalized the kinetic equations to units in which
$D=\Lambda=k_BT=1$. Redimensionalizing them,
taking the typical values $D=10^{-7}-10^{-8}$cm$^2$/s (for lipids
in a liquid-phase membrane),
$\mu = 1$cp ($H_2O$ at $20^{\circ}$C) and the typical unstable modes
$q \approx q_c=0.5$, we find that the size of the patterns obtained
above lies within the range $2-20 \mu$m,
which is accessible in giant vesicle ($10-100 \mu$m in diameter) experiments.

\section{Membranes of miscible components: Linearization and shape fluctuations}
\label{stable}

In this section we consider a parameter region where the membrane is thermodinamically
stable. The instability in the previous section
was caused by the immiscibility of the two lipid components, while
we now consider the case of miscible membrane constituents.
Notice that the two situations could correspond to the same physical
system but at different temperatures, below (unstable)
or above (stable) the critical temperature. For the stable case
we adopt the following free energy functional,
\begin{equation}
{\mathcal{F}}= \int \left[ -\frac{\alpha}{2} \phi^{2} + \frac{\kappa}{2}
(\nabla^2 h - \phi H_0)^2 + \frac{\sigma}{2}
\left( |{\mathbf{\nabla}} h|\right) ^{2}\right] dx dy,
\label{fgenstab}
\end{equation}
where $\alpha$ is now negative, and the nonlinear ($\beta$) and
line tension ($\gamma$) terms have been removed
since they are irrelevant in the absence of phase segregation. An
additional surface tension term has been included
in order to study membranes under tension.

In the equilibrium situation (no reaction), the concentration
field can be integrated out ($\frac{\partial {\mathcal{F}}}{\partial \phi}=0$)
from Eq.~(\ref{fgenstab}), leading to $\phi_{eq}=\frac{\kappa H_0}
{\kappa H_0^2 -\alpha}\nabla^2 h$. Inserting $\phi_{eq}$
in Eq.~(\ref{fgenstab}) we obtain an effective free energy for $h$ alone,
\begin{equation}
{\mathcal{F}}_{eff}= \int \left[
\frac{\kappa_{eff}}{2} (\nabla^2 h)^2 + \frac{\sigma}{2}
\left( |{\mathbf{\nabla}} h|\right) ^{2}\right] dx dy,
\label{fgenstab2}
\end{equation}
with an effective rigidity modulus \cite{andel}
\begin{equation}
\kappa_{eff}=\kappa \left[ 1 - \left( \frac{\kappa H_0^2}
{\kappa H_0^2 -\alpha} \right) \right].
\label{keff}
\end{equation}
Notice that, as a consequence of the composition/curvature coupling,
the different preferred curvature of the lipid components acts
to reduce the rigidity modulus from $\kappa$ to
$\kappa_{eff}$. Note also that $\kappa_{eff} > 0$ for negative values
of $\alpha$, so that the stability of the membrane is assured
above the critical temperature.

The dynamics for the immiscibility situation has so far been considered
to be strictly deterministic. Here, however,
in order to obtain dynamical steady state functions, we
add stochastic forces to the kinetic equations
and perform an average over an appropriate ensemble.
The dimensionless kinetic equation for $\phi$, once the reactive
terms are included, reads
\begin{equation}
\frac{\partial \phi}{\partial t} = 
\nabla^{2} \left[ \frac{\partial {\mathcal{F}}}{\partial \phi}
\right] - \Gamma (\phi - \phi_0) + {\mathbf{\nabla}}\cdot {\bf f}_{\phi},
\label{phivarstab}
\end{equation}
where the last term corresponds to a conserving Gaussian noise with
correlations $\langle f_{\phi}({\bf r},t) f_{\phi}({\bf r'},t')\rangle =
2 k_BT \delta ({\bf r}-{\bf r'}) \delta(t-t')$.
For the curvature order parameter we again adopt
the relaxational equation for a permeable surface,
\begin{equation}
\frac{\partial h }{\partial t}=- \Lambda \frac{\delta {\mathcal{F}}}{\delta
h } + f_h,
\label{hvarstab}
\end{equation}
with $\Lambda =1$ in dimensionless units, and
$f_h$ is a thermal equilibrium noise with correlations
$\langle f_h({\bf r},t) f_h({\bf r'},t')\rangle = 2 k_BT \delta ({\bf r}-{\bf r'}) \delta(t-t')$.

The kinetic equations become
\begin{align}
\begin{split}
\frac{\partial \phi}{\partial t} = 
(\kappa H_0^{2} - \alpha) \nabla^{2} \phi -\kappa H_0 \nabla^{4}h
 - \Gamma \left( \phi - \phi_0 \right)
+ {\mathbf{\nabla}}\cdot {\bf f}_{\phi} \\
\frac{\partial h }{\partial t} = - \kappa \nabla^{4}h -\sigma \nabla^{2} h
+\kappa H_0 \nabla^{2} \phi + f_h.
\label{kineqstab}
\end{split}
\end{align}

The important quantity to characterize membrane shape fluctuations
is the height variance $\langle |h_q|^{2} \rangle$ at wave number $q$, which
is calculated by linearizing the kinetic equations (\ref{kineqstab})
and solving them in Fourier space.
If $\hat{\phi}({\bf q},\omega )$ and $\hat{h}({\bf q},\omega )$ are
the Fourier transforms of
$\phi({\bf r},t)$ and $h({\bf r},t)$, respectively,
Eqs.~(\ref{kineqstab}) can be written as
\begin{equation}
i \omega \left(
\begin{tabular}{c}
$\hat{\phi}({\bf q},\omega)$  \\
$\hat{h}({\bf q},\omega)$
\end{tabular}
\right) = \left(
\begin{tabular}{cc}
$a_{11}$ & $a_{12}$ \\
$a_{21}$ & $a_{22}$
\end{tabular}
\right) \times \left(
\begin{tabular}{c}
$\hat{\phi}({\bf q},\omega)$ \\
$\hat{h}({\bf q},\omega)$
\end{tabular}
\right) + \left(
\begin{tabular}{c}
$i {\bf q} \cdot {\bf f}_{\phi}$  \\
$f_h$
\end{tabular}
\right),
\label{2x2}
\end{equation}
where
\begin{align}
\begin{split}
a_{11}&=-\left( \kappa H_0^{2} - \alpha \right) q^{2} - \Gamma \\
a_{12}&=-\kappa H_0 q^{4}  \\
a_{21}&=-\kappa H_0 q^{2}  \\
a_{22}&=-\kappa q^{4} + \sigma q^{2} \, .
\label{coefic}
\end{split}
\end{align}
Solving these coupled equations, and using the statistical properties of the thermal noises, yields
\begin{equation}
\langle \hat{h}(q,\omega) \hat{h}^{*}(q,\omega) \rangle =
\frac{2 k_BT \left(a_{21}^{2} q^{2} + \omega^{2}  +a_{11}^{2} \right) }
{\omega^{2} \left( a_{11} + a_{22} \right)^{2}+\left( a_{12} a_{21}
- a_{11} a_{22} + \omega^{2} \right)^{2}}.
\label{hqhq}
\end{equation}
On integrating over $\omega$, the height variance follows the general
expression,
\begin{equation}
\langle |h_q|^{2} \rangle = \frac{1}{\nu-\eta} \left[ \frac{A}{\sqrt{\nu \eta}} 
+ A' \right] \left[ \sqrt{-\eta} - \sqrt{-\nu} \right],
\label{hq2}
\end{equation}
where $\nu$ and $\eta$ are negative variables that read 
\begin{eqnarray}
\left( \begin{tabular}{c}
$\nu$ \\
$\eta$
\end{tabular} \right)
= -\frac{B}{2} \pm \frac{1}{2} \sqrt{B^{2}-4C},
\label{nuieta}
\end{eqnarray}
and
\begin{align}
\begin{split}
A&=2 k_BT \left( a_{21}^{2} q^{2} + a_{11}^{2} \right) \\
A'&=2 k_BT \\
B&=a_{11}^{2}+a_{22}^{2}+2a_{21}a_{12} \\
C&=\left( a_{11}a_{22}-a_{21}a_{12} \right)^{2}.
\label{coefa}
\end{split}
\end{align}

In the absence of the reaction, the evaluation of Eq.~(\ref{hq2}) in the
long-wavelength limit for a tensionless membrane
leads to $\langle |h_q|^{2} \rangle = k_BT/\kappa_{eff} q^{4}$. This
behavior is truncated and replaced
by $k_BT/\sigma q^{2}$ for a membrane under tension if only
the dominant terms at small $q$ are retained.
Keeping the dominant and the first subdominant terms, one recovers
the well known expression for the height variance
in an equilibrium membrane,
\begin{equation}
\langle |h_q|^{2} \rangle = \frac{k_BT}{\sigma q^{2} + \kappa_{eff} q^{4}}.
\label{hq2eq}
\end{equation}
Notice that this result is obtained in a much simpler way from an
equilibrium average
using the effective free energy in Eq.~(\ref{fgenstab2}).

One of the usual experiments to study equilibrium membrane shape
fluctuations is based on the micropipet
aspiration technique~\cite{evans}. In these experiments, a pressure
difference is applied inside a micropipet in contact with a
vesicle membrane. This creates a tension in the membrane that pulls
the excess area $\Delta S$ due to the
thermal shape fluctuations inside the micropipet. 
By means of this technique, the areal strain $\bar{\alpha} \equiv \Delta S / S$
is obtained experimentally for different values of the applied tension $\sigma$.
According to Eq.~(\ref{hq2eq}), the relative areal strain 
$\Delta \bar{\alpha} \equiv \bar{\alpha}
- \bar{\alpha}_0$ ($\bar{\alpha}_0$ being the areal
strain for a reference value $\sigma_0$) can be
calculated analytically~\cite{nuovo,fournier}:
\begin{equation}
\Delta \bar{\alpha} = \frac{k_BT}{8\pi\kappa_{eff}}
\ln \left( \frac{\sigma}{\sigma_0} \right).
\label{areal3}
\end{equation}
Therefore, the slope of the logarithm of $\sigma$ versus
$\Delta \bar{\alpha}$ obtained by the micropipet technique yields
a measure of the effective bending modulus. Note that in these
experiments a certain tension $\sigma$ is needed, but it has to
be small since otherwise the $\kappa_{eff} q^{4}$ term in
Eq.~(\ref{hq2eq}) would be insignificant compared to $\sigma q^{2}$.

Now the question is, how is the shape fluctuation spectrum
affected by the presence of the reaction? In other words,
how would the nonequilibrium membranes described here behave in
micropipet experiments? To answer this, we
evaluate Eq.~(\ref{hq2}) for $\Gamma \neq 0$ and keep
only the dominant terms in the limit $q \rightarrow 0$. For tensionless
membranes we get $\langle |h_q|^{2} \rangle = k_BT/\kappa q^{4}$
and for membranes under tension
we recover $\langle |h_q|^{2} \rangle = k_BT/\sigma q^{2}$.
Keeping the dominant terms for both limits, in the weak
tension regime the fluctuation spectrum reads
\begin{equation}
\langle |h_q|^{2} \rangle = \frac{k_BT}{\sigma q^{2} + \kappa q^{4}}.
\label{hq2noeq}
\end{equation} 
Comparing this result with the fluctuation spectrum obtained
for membranes with active proteins~\cite{prost3,sum,prost4},
no novel nonequilibrium contribution to the height fluctuations is
found here. The effect of the nonequilibrium reaction in stable
membranes results in a change between a regime governed by
$\kappa_{eff}$ to another one with the actual rigidity $\kappa$.
Thus, the reaction makes the membrane more rigid by simply removing
the composition/curvature coupling effect that diminished the
rigidity in the equilibrium situation. This change is
more evident when the two components of the bilayer are rather
different in shape and miscible but close to the critical
temperature. In this situation (large
$H_0$ and $\bar{\alpha} \lesssim 0$) the difference between
$\kappa$ and $\kappa_{eff}$ might be experimentally observable.

\begin{figure}[h!]
\begin{center}
\includegraphics[width=8cm ]{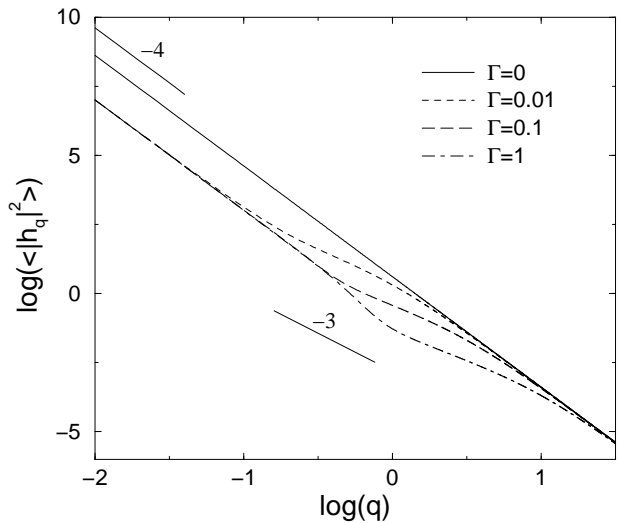}
\caption{Log-log plot for $\langle |h_q|^{2} \rangle$ evaluated
from Eq.~(\ref{hq2}) in a tensionless membrane for different
values of $\Gamma$. The other parameters are $\alpha=-0.01$,
$\kappa=10$ and $H_0=0.2$ ($\kappa_{eff}=2$ in these cases). The slopes
$-1/4$ and $-1/3$ are plotted to identify the different regimes.}
\label{fighq2}
\end{center}
\end{figure}

Some caution, however, must be observed when deriving Eq.~(\ref{hq2noeq}).
In order to evaluate Eq.~(\ref{hq2}) for small $q$, we
considered the terms $\Gamma^{2}$ to be dominant with respect to
the terms $2\Gamma \left(\kappa H_0^{2} - \alpha \right) q^{2}$
(subdominant) and also with respect to
the subsubdominant quartic contributions
$\left( \sigma + \left(\kappa H_0^{2} - \alpha \right) \right) q^{4}$
(which are those that
led to Eq.~(\ref{hq2eq})). However, the analysis becomes more difficult
if one notices that we are not in the region of asymptotically
small wavenumbers since the smallest $q$ accessible to an experiment is of
order $1/L$, where $L$ is the linear system size.
Thus, one must look in detail at the values of the parameters in order
to determine exactly which terms initially considered
subdominant might indeed be dominant when the reaction is present.
$\Gamma ^{2}$ is dominant if
$\Gamma > 2 \left(\kappa H_0^{2} - \alpha \right)
q_{min}^{2}$, with $q_{min} \sim 1/L$ and thus dependent on the
system size.  We know that for a sufficiently strong reactive
process, $\Gamma^{2}$ dominates and Eq.~(\ref{hq2noeq}) holds, but for
intermediate values of $\Gamma$ the spectrum could change significatively.
When $2\Gamma \left(\kappa H_0^{2} - \alpha \right) q^{2}$
becomes dominant, $\langle |h_q|^{2} \rangle = \frac{k_BT}
{\left( \Gamma \kappa / \sqrt{8\Gamma
\left(\kappa H_0^{2} - \alpha \right)}\right) q^{3}}$ for a
tensionless membrane. These
regimes are captured in Fig.~\ref{fighq2}, where a set of $\langle |h_q|^{2} \rangle$ curves is plotted for $\sigma=0$.
Very large and zero values for $\Gamma$ show the $q^{-4}$ behavior,
although with different rigidity factors ($\kappa$ and
$\kappa_{eff}$, respectively). At intermediate
$\Gamma$, the $q^{-3}$ behavior appears as predicted above.

The inclusion of hydrodynamic interactions can again be accomplished
by considering $\Lambda = \left( 4 \mu q \right)^{-1}$
in the Fourier transform of Eq.~(\ref{hvarstab}). This modifies
the coefficients $a_{21}$ and $a_{22}$ in Eq.~(\ref{coefic}), which
are now divided by $\left( 4 \mu q \right)$. Accordingly, the
$f_h$ thermal noise correlations also change to 
$\langle f_h({\bf r},t) f_h({\bf r'},t')\rangle =
\frac{2 k_BT}{\left( 4 \mu q \right)} \delta ({\bf r}-{\bf r'})
\delta(t-t')$, and therefore the parameter $A'$ in Eq.~(\ref{coefa})
has to be divided by $\left( 4 \mu q \right)$ as well. However,
with these modifications the evaluation of $\langle |h_q|^{2} \rangle$
again leads to the results in Eqs.~(\ref{hq2eq}) and (\ref{hq2noeq}).

\section{Conclusions}
\label{concl}

Starting with a simple model of a deformable reactive membrane
composed of two differently shaped molecules, we show that
stationary finite-sized patterns may appear under some parameter
conditions for the immiscibility situation
as a result of the competition between phase segregation and
reaction. These structures involve 
heterogeneous distributions of composition and curvature
whose sizes are determined by the nonequilibrium reactive process.
For typical values of the viscosity of water and lipid lateral
diffusion constants in bilayers, and at normal room temperatures,
such patterns are predicted to have a size of a few microns (see
the discussion at the end of Sec.~\ref{numres}).
Therefore, this behavior would correspond to a reliable pattern formation
mechanism in lipid membranes which we believe to be experimentally accessible
in giant synthetic vesicles.
The amplitude of these patterns is modulated by the bilayer rigidity
and the spontaneous curvature of its components. In our numerical
calculations we have used realistic typical values for the rigidity,
while the spontaneous curvature depends on the specific geometry of the
membrane constituents. We specifically propose that
azobenzene compounds, which are known to show amphiphilic behavior in
Langmuir monolayers, and whose shapes are strongly modified by means
of well-known photoisomerization reactions~\cite{rau,our}, might be suitable to
test our predictions. The selection of the applied light wavelength and its intensity
may determine both the fraction of the two isomers ($\phi_0$) and the
strength of the reactive process ($\Gamma$), respectively. 
Thus, experimental work on synthetic vesicle membranes made of these compounds
can be specifically designed to confirm the results of this model.

In the same context, for miscible components membranes we have found a
difference between the equilibrium and nonequilibrium situations.
The effects of the
composition/curvature coupling and the reactive process
on the membrane rigidity are established. Micropipet experiments
with the proposed membrane systems might confirm these results.

\section*{Acknowledgments}

This work was supported by the $Ajuts\, Gaspar\, de\, Portol\grave{a}$
program from DURSI (Generalitat de Catalunya). Two of us (R.R and J.B.)
greatfully acknowledge the $Ram\acute{o}n\, y\, Cajal$ program
(Ministerio de Ciencia y Tecnolog\'{\i}a) that provides
their researcher contracts. Partial support was provided by the Office
of Basic Energy Sciences at the U. S. Department of Energy
under Grant No. DE-FG03-86ER13606. The authors
gratefully thank Dr. A. J. Pons for helpful discussions on the
amplitude equations derivation.

\appendix

\section{Derivation of the amplitude equations}

The starting point of the analysis is the one dimensional version of the model
for reactive membranes presented herein,%

\begin{align}
\varphi_{t}  &  =\left(  \kappa H_{0}^{2}-\alpha\right)  \varphi_{xx}%
+3\beta\left(  \varphi+\phi_{0}\right)^{2}  \varphi_{xx}+6\beta\left(
\varphi+\phi_{0}\right)  \left(  \varphi_{x}\right)  ^{2}-\gamma\varphi
_{xxxx}\nonumber\\
&  -\kappa H_{0}h_{xxxx}-\frac{\left(  \alpha-3\beta\phi_{0}^{2}\right)  ^{2}%
}{4\gamma}\left(  1-\varepsilon\right)  \varphi\text{,}\label{eq}\\
h_{t}  &  =-\kappa h_{xxxx}+\kappa H_{0}\varphi_{xx}\text{.}\nonumber
\end{align}

For convenience, we have introduced in Eqs. (\ref{eq}) some notation changes
and definitions with respect to the ones that appear in Eqs. (\ref{kineq}).
Thus, in
Eqs. (\ref{eq}) subscripts indicate partial derivatives, the field
$\varphi=\phi-\phi_{0}$ has been defined, and we have introduced the control
parameter $\varepsilon=\left(  \Gamma_{c}-\Gamma\right)  /\Gamma_{c}$ that
accounts for the \textquotedblleft distance\textquotedblright\ to the
bifurcation between a homogenous state ($\varepsilon<0$) and pattern
formation ($\varepsilon>0$). The homogeneous state according to this
definitions corresponds to $\varphi=0$ and arbitrary $h=\overline{h}$.

As shown in Sec. (\ref{modan}), by linearizing Eqs. (\ref{eq}) one can easily check
that if $\varepsilon>0$ then the homogeneous state becomes unstable and,%
\begin{align}
\begin{split}
\varphi\left(  x\right)   &  =A\exp\left(  iq_{c}x\right)  +A^{\ast}%
\exp\left(  -iq_{c}x\right)  \text{,}\\
h\left(  x\right)   &  =B\exp\left(  iq_{c}x\right)  +B^{\ast}\exp\left(
-iq_{c}x\right)  \text{,}
\label{2}
\end{split}
\end{align}
is a solution in the steady state if $q_{c}^{2}=\left(  \alpha-3\beta\phi
_{0}^{2}\right)  /\left(  2\gamma\right)  $.
It is worth noting that in Eqs. (\ref{2}) we have arbitrarily taken $\overline{h}=0$
without any loss of generality. Moreover, by substituting
(\ref{2}) into (\ref{eq}) and expanding up to the first harmonic,
$\mathcal{O}\left(  \exp\left(  iq_{c}x\right)  \right)  $, one finds that the
amplitudes scale as a function of $\varepsilon$ as $A,B\sim\sqrt{\varepsilon}%
$. Thus, we expect that near the bifurcation the following expansion holds,%
\begin{align}
\begin{split}
\varphi &  =\sum_{n=1}^{\infty}\varepsilon^{n/2}\varphi^{\left(  n\right)
}\text{,}\\
h &  =\sum_{n=1}^{\infty}\varepsilon^{n/2}h^{\left(  n\right)  }%
\text{.}
\label{4}
\end{split}
\end{align}

By computing the linear growth rate, $\omega_q$, that is, the largest
eigenvalue of the linear problem (see Eqs. (\ref{coefl})), we also note that,%
\[
\left.  \omega_{q}\right\vert _{\substack{q\rightarrow q_{c}\\\varepsilon\rightarrow 0}}\simeq
C_{1}\varepsilon+C_{2}\left(  q-q_{c}\right)  ^{2}\text{,}%
\]
where $C_{i}$ are constants. Thus, as a function of $\varepsilon$ the width of
the band of unstable modes scales as $\sim\varepsilon^{1/2}$. Then, since
\emph{all} modes $\exp\left(  iqx\right)  $ can be written as $\exp\left(
i\left(  q-q_{c}\right)  x\right)  \exp\left(  -iq_{c}x\right)  $, a
\emph{separation of spatial scales} can be performed between the most unstable
mode (fast) and the rest of the modes of the unstable band (slow). Let us call
the slow modulation spatial scale $X$, such that $X=\varepsilon^{1/2}x$, where
$x$ will now stand for the fast spatial scale. We also note
that
\[
\left.  \exp\left(  \omega_{q}t\right)  \right\vert _{\substack{q\rightarrow q_{c}%
\\\varepsilon\rightarrow 0}}\simeq\exp\left(  \varepsilon t\right)  \text{.}%
\]
Therefore we can define a slow time scale as a function of the control
parameter, $T=\varepsilon t$.  The separation of scales can be implemented in
Eqs. (\ref{eq}) by replacing the spatial and temporal derivatives according to
the chain rule such that $\partial_{x}\rightarrow\partial_{x}+\varepsilon
^{1/2}\partial_{X}$ and $\partial_{t}\rightarrow\varepsilon\partial_{T}$.

By implementing the separation of scales and substituting Eqs. (\ref{4}) into
Eqs. (\ref{eq}) we obtain a rather cumbersome expansion in terms of
$\varepsilon$. The lowest order contribution is of $O\left( \varepsilon^{1/2} \right)$
and reads%
\begin{align}
\begin{split}
\left(  3\beta\phi_{0}^{2}+H_{0}^{2}\kappa-\alpha\right)  \varphi
_{xx}^{\left(  1\right)  }-\gamma\varphi_{xxxx}^{\left(  1\right)  }%
-\frac{\left(  \alpha-3\beta\phi_{0}^{2}\right)  ^{2}}{4\gamma}\varphi
^{\left(  1\right)  }-H_{0}\kappa h_{xxxx}^{\left(  1\right)  } &
=0\text{,}\\
H_{0}\kappa\varphi_{xx}^{\left(  1\right)  }-\kappa h_{xxxx}^{\left(
1\right)  } &  =0\text{.}
\label{7}
\end{split}
\end{align}
Note that Eqs. (\ref{7}) correspond to the linearized version of
Eqs. (\ref{eq}) in the stationary state. We define the linear operator%
\[
\mathbb{L}\equiv\left(
\begin{array}
[c]{cc}%
-\frac{\left(  \alpha-3\beta\phi_{0}^{2}\right)  ^{2}}{4\gamma}+\left(
3\beta\phi_{0}^{2}+H_{0}^{2}\kappa-\alpha\right)  \partial_{xx}-\gamma
\partial_{xxxx} & -H_{0}\kappa\partial_{xxxx}\\
H_{0}\kappa\partial_{xx} & -\kappa\partial_{xxxx}%
\end{array}
\right)  \text{.}%
\]
Then, Eqs. (\ref{7}) can be trivially written as $\mathbb{L}\mathbf{\chi}%
_{1}=0$, where $\left(  \mathbf{\chi}_{n}\right)  ^{T}=\left(  \varphi
^{\left(  n\right)  },h^{\left(  n\right)  }\right)  $. The contributions of
the next order,
$O \left(\varepsilon\right)$, are $\mathbb{L}\mathbf{\chi}_{2}=\mathbf{\psi}_{2}\left(
\left\{  \varphi^{\left(  1\right)  };h^{\left(  1\right)  }\right\}  \right)
$ where $\mathbf{\psi}_{2}^{T}=\left(  \mathbf{\psi}_{2}^{\left(  a\right)
}\mathbf{,\psi}_{2}^{\left(  b\right)  }\right)  $,%
\begin{align*}
\mathbf{\psi}_{2}^{\left(  a\right)  } &  =-6\beta\phi_{0}\left(  \left(
\varphi_{x}^{\left(  1\right)  }\right)  ^{2}+\varphi^{\left(  1\right)
}\varphi_{xx}^{\left(  1\right)  }\right)  +2\left(
\alpha-3\beta\phi_{0}%
^{2}-H_{0}^{2}\kappa\right)  \varphi_{xX}^{\left(  1\right)  }+\\
&  +4\gamma\varphi_{xxxX}^{\left(  1\right)  }+4H_{0}\kappa
h_{xxxX}^{\left(
1\right)  }\text{,}\\
\mathbf{\psi}_{2}^{\left(  b\right)  } &  =2\kappa\left(  -H_{0}\varphi
_{xX}^{\left(  1\right)  }+2h_{xxxX}^{\left(  1\right)  }\right)
\text{,}%
\end{align*}

Finally, at order $\varepsilon^{3/2}$ we get $\mathbb{L}\mathbf{\chi}%
_{3}=\mathbf{\psi}_{3}\left(  \left\{  \varphi^{\left(  1\right)  }%
,\varphi^{\left(  2\right)  };h^{\left(  1\right)  },h^{\left(  2\right)
}\right\}  \right)  $, where $\mathbf{\psi}_{3}^{T}=\left(  \mathbf{\psi}%
_{3}^{\left(  a\right)  }\mathbf{,\psi}_{3}^{\left(  b\right)  }\right)  $ reads,%

\begin{align*}
\mathbf{\psi}_{3}^{\left(  a\right)  } &  =\varphi_{T}^{\left(  1\right)
}-\Gamma_{c}\varphi^{\left(  1\right)  }-6\beta\varphi_{x}^{\left(
1\right)
}\left(  \varphi^{\left(  1\right)  }\varphi_{x}^{\left(  1\right)  }%
+2\phi_{0}\left(  \varphi_{X}^{\left(  1\right)  }+\varphi_{x}^{\left(
2\right)  }\right)  \right)  +\\
&  +\left(  \alpha-H_{0}^{2}\kappa\right)  \left(  \varphi_{XX}^{\left(
1\right)  }+2\varphi_{xX}^{\left(  2\right)  }\right)  -3\beta\left[
\phi
_{0}^{2}\left(  \varphi_{XX}^{\left(  1\right)  }+2\varphi_{xX}^{\left(
2\right)  }\right)  +\right.  \\
&  \left.  +\varphi_{xx}^{\left(  1\right)  }\left(  \left(
\varphi^{\left(
1\right)  }\right)  ^{2}+2\phi_{0}\varphi^{\left(  2\right)  }\right)
+2\phi_{0}\varphi^{\left(  1\right)  }\left(  2\varphi_{xX}^{\left(
1\right)
}+\varphi_{xx}^{\left(  2\right)  }\right)  \right]  +\\
&  +2\gamma\left(  3\varphi_{xxXX}^{\left(  1\right)  }+2\varphi
_{xxxX}^{\left(  2\right)  }\right)  +2H_{0}\kappa\left(
3h_{xxXX}^{\left(
1\right)  }+2h_{xxxX}^{\left(  2\right)  }\right)  \text{,}\\
\mathbf{\psi}_{3}^{\left(  b\right)  } &  =h_{T}^{\left(  1\right)  }%
+\kappa\left(  2\left(  3h_{xxXX}^{\left(  1\right)  }+2h_{xxxX}^{\left(
2\right)  }\right)  -H_{0}\left(  \varphi_{XX}^{\left(  1\right)  }%
+2\varphi_{xX}^{\left(  2\right)  }\right)  \right)  \text{.}%
\end{align*}

We could continue up to any order with the expansion. In all cases we will
obtain a nonlinear equation, such that at order $\varepsilon^{n/2}$,%
\begin{equation}
\mathbb{L}\mathbf{\chi}_{n}=\mathbf{\psi}_{n}\left(  \left\{  \varphi^{\left(
1\right)  },\ldots,\varphi^{\left(  n-1\right)  };h^{\left(  1\right)
},\ldots,h^{\left(  n-1\right)  }\right\}  \right)  \text{.}\label{ecuaciones}%
\end{equation}
However, at order $\varepsilon^{3/2}$ we are already able to extract a closed
evolution equation for the amplitudes of the pattern and so we will stop at that order.

Our task is to solve the hierarchy of equations given by (\ref{ecuaciones}).
At order $\varepsilon^{1/2}$ the problem is homogeneous and with appropriate
boundary conditions,%
\begin{align}
\begin{split}
\varphi^{\left(  1\right)  } &  =A\left(  X,T\right)  \exp\left(
iq_{c}x\right)  +A^{\ast}\left(  X,T\right)  \exp\left(  -iq_{c}x\right)
\text{,}\\
h^{\left(  1\right)  } &  =B\left(  X,T\right)  \exp\left(  iq_{c}x\right)
+B^{\ast}\left(  X,T\right)  \exp\left(  -iq_{c}x\right)  \text{,}
\label{12}
\end{split}
\end{align}
is a solution. However, the amplitudes $A$ and $B$ are undetermined at this
point. The subsequent orders are no longer homogeneous and therefore their
solvability can not be ensured unless one implements the
so-called \emph{Fredholm alternative theorem} \cite{newell}. In our case the
application of the theorem simply states, as a recipe, that for
Eqs. (\ref{ecuaciones}) to have a solution the functions $\psi_{n}$\ can not
contain the fundamental mode $\exp\left(  \pm iq_{c}x\right)  $. Thus, by
substituting the solution (\ref{12}) into the next order or the hierarchy and
imposing the solvability condition we obtain%
\begin{align*}
\varphi^{\left(  2\right)  } &  =-\frac{8\beta\phi_{0}}{3\left(  \alpha
-3\beta\phi_{0}^{2}\right)  }\left[  \left(  A\left(  X,T\right)  \right)
^{2}\exp\left(  i2q_{c}x\right)  +\left(  A\left(  X,T\right)  ^{\ast}\right)
^{2}\exp\left(  -i2q_{c}x\right)  \right]  \text{,}\\
h^{\left(  2\right)  } &  =\frac{4\beta\phi_{0}\gamma H_{0}}{3\left(
\alpha-3\beta\phi_{0}^{2}\right)  ^{2}}\left[  \left(  A\left(  X,T\right)
\right)  ^{2}\exp\left(  i2q_{c}x\right)  +\left(  A\left(  X,T\right)
^{\ast}\right)  ^{2}\exp\left(  -i2q_{c}x\right)  \right]  \text{.}%
\end{align*}
Once again, the value of $A$ and $B$ can not be determined at this order.
However, at order $\varepsilon^{3/2}$ the application of the Fredholm theorem
provides the conditions that determine the values of the amplitudes $A$, $B$.
These conditions constitute the amplitude equations for our pattern forming
system,
\begin{align}
\nonumber \partial_{T}A &  =\frac{\left(  \alpha-3\beta\phi_{0}^{2}\right)  ^{2}%
}{4\gamma}A-\beta\frac{\left(  3\alpha+7\beta\phi_{0}^{2}\right)  }{2\gamma
}A\left\vert A\right\vert ^{2}+\label{ae2}\\
&  +2\left(  \alpha-3\beta\phi_{0}^{2}+\frac{H_{0}^{2}\kappa}{2}\right)
\partial_{XX}A+\frac{3H_{0}\kappa\left(  \alpha-3\beta\phi_{0}^{2}\right)
}{\gamma}\partial_{XX}B\text{,} \\
\partial_{T}B &  =H_{0}\kappa\partial_{XX}A+\frac{3\kappa\left(  \alpha
-3\beta\phi_{0}^{2}\right)  }{\gamma}\partial_{XX}B\text{.}\nonumber
\end{align}

Equations (\ref{ae2}) can be rewritten in terms of $x$ and $t$ to readily
obtain Eqs. (\ref{ae}).


\end{document}